\documentclass{article}

\usepackage{graphicx}
\usepackage{psfig}
\usepackage{epsfig}
\usepackage[round]{natbib}

\title{A note on the use of the word 'likelihood' in statistics and meteorology}

\author{Stephen Jewson, Anders Brix and Christine Ziehmann
\footnote{\emph{Correspondence address}: RMS, 10 Eastcheap,
London, EC3M 1AJ, UK. Email: \texttt{x@stephenjewson.com}}\\
Risk Management Solutions, London, United Kingdom}

\begin{document}

\maketitle

\begin{abstract}
We highlight the different uses of the word \emph{likelihood} that have arisen in statistics and
meteorology, and make the recommendation that one of these uses should be dropped to prevent confusion
and misunderstanding.
\end{abstract}

\section{Introduction}

We discuss the different meanings of the word \emph{likelihood} as used in classical statistics and
meteorology. In section~\ref{statistics} we describe how the word is used in classical
statistics, and in section~\ref{meteorology} we describe how it has been used in meteorology.
In section~\ref{discussion} we discuss the differences and express the opinion that one of the
two uses should be dropped. In section~\ref{conclusions} we summarize.

\section{Classical statistics definition of likelihood}
\label{statistics}

Imagine that we have two datasets $x_i$ and $y_i$, for a range of values of $i$.
We might attempt to build a statistical model for $x_i$ in terms of $y_i$, or $y_i$ in terms of $x_i$.

As an example we will consider linear regression, and will consider building a model for
$y_i$ in terms of $x_i$. The model we will consider can be written in the following equivalent ways:

\begin{equation}
  y_i=\alpha+\beta x_i+\sigma e_i
\end{equation}

where 

\begin{equation}
 e_i \sim N (0,1)
\end{equation}

or

\begin{equation}
  y_i \sim N (\alpha+\beta x_i, \sigma^2)
\end{equation}

or 

\begin{equation}
  p(y_i | x_i) = N ( \alpha+\beta x_i, \sigma^2)
\end{equation}

To fit the parameters $(\alpha,\beta,\sigma)$
of this model given the data $x_i$ and $y_i$ 
one would typically consider finding those parameters 
that maximise the likelihood $L(\alpha,\beta,\sigma)$, where $L$ is defined as:

\begin{equation}
  L(\alpha,\beta,\sigma)
           = p(y | x,\alpha,\beta,\sigma) 
\end{equation}

In the case of linear regression, this happens to be equivalent to finding parameters that minimise
the sum of squared errors. 

We note that the definition of $L$ as $L(\alpha,\beta,\sigma)=p(y|x,\alpha,\beta,\sigma)$ 
arises because we are using $x_i$ to model $y_i$ i.e. $x$ is the 'input', 
'independent variable', 'predictor', 'regressor', 'explanatory variable' or 'covariate', 
while $y$ is the 'output', 'dependent variable', 'regressand', 'response variable' or 'predictand'.

If we had set out to use $y_i$ to model $x_i$ (using linear regression, or any other model)
then the likelihood would have been defined as $L(\theta)=p(x|y, \theta)$ where $\theta$ represents
the parameters of this new model. 
We see that the definition of likelihood is thus entirely dependent on what is being used to model what, and
what model is being used. 

In general, once we have decided on a model then the likelihood is the probability density 
(or, for discrete rather than continuous models, just the probability) of the predictand given
the predictors, as a function of the parameters of the model. 

Use of the word likelihood in this context comes from the original works of Fisher in the 1920s, 
such as~\citet{fisher1922}. A recent overview of the use of the likelihood in statistical inference is 
given by~\citet{casellab02}, and likelihood-based fitting of parameters is also discussed in~\citet{nr}.

\subsection{Application to making probabilistic forecasts}
\label{application}

We now consider a meteorological application: making probabilistic forecasts.
Probabilistic forecasts are made by taking the inherently non-probabilistic output from numerical models 
(consisting of, for example, single
integrations, ensemble members, or ensemble means and spreads) and fitting a statistical model to them to generate
probabilities. In other words, given a non-probabilistic model forecast $f$ we make
a probabilistic forecast of the observations $o$. Our forecast can be written as $p(o|f)$ i.e. a probability
distribution of different possible observations, given the forecast we have available. Note that the notation
$p(o|f)$ does not yet specify what model is used to convert $f$ to $o$. 

For temperature, a reasonable way to make a probabilistic forecast is to use linear regression, with the input
$f_i$ being either a single forecast or an ensemble mean. This then means we can write:

\begin{equation}
  p(o_i | f_i)=N ( \alpha+\beta f_i, \sigma^2)
\end{equation}

If we wish to use information from the mean $m_i$ and spread $s_i$ of an ensemble forecast, then
we can use the spread regression model of~\citet{jewsonbz03a}:

\begin{equation}
  p(o_i | f_i)=N ( \alpha+\beta m_i,(\gamma+\delta s_i)^2)
\end{equation}

The usual way to fit the parameters of either of these models would be to 
find those parameters that maximise the
likelihood, defined as $L=L(\alpha,\beta,\sigma)=p(o|f,\alpha,\beta,\sigma)$ (for the regression model)
or $L=L(\alpha,\beta,\gamma,\delta)=p(o|f,\alpha,\beta,\gamma,\delta)$ (for the spread regression model).
The likelihood is defined as $p(o|f)$ simply because we are trying to predict
the observations $o$ from the forecast $f$. If, for some reason, we wanted to predict the forecast from the 
observations (it is not immediately obvious why one would want to do this, but there may be reasons), then we would
define the likelihood as $p(f|o)$.

\section{Murphy and Winkler definition of likelihood} 
\label{meteorology}

\citet{murphyw87} (henceforth MW) discuss ways in which one can validate probabilistic forecasts, 
and in particular introduce the 
following definitions:

\begin{itemize}
  \item $p(f|o)$ is the \emph{likelihood}
  \item $p(o)$ is the \emph{base rate}
  \item $p(o|f)$ is the \emph{calibration}
  \item $p(f)$ is the \emph{refinement}
\end{itemize}

Following this paper, a number of other meteorologists 
(such as~\citet{jolliffes03} and~\citet{wilks95}) have used the word \emph{likelihood} to refer to $p(f|o)$ and
the word \emph{calibration} to refer to $p(o|f)$.

\section{Discussion}
\label{discussion}

We see that the MW definition of the word likelihood is subtly different from the
original definition as used in classical statistics. 
In particular, MW define likelihood
once and for all as $p(f|o)$ irrespective of whether $o$ is being modelled in terms of $f$, or $f$ is being 
modelled in terms of $o$. The classical statistics definition of likelihood, 
on the other hand, depends on what is being used to model what.

This creates some confusion,
especially when one tries to apply classical statistical methods to forecast calibration as described
in section~\ref{application}.
Because of this, we advocate that the MW definition should
not be used.
Our reasons for taking this position are:

\begin{itemize}
  \item The classical statistical definition of the word likelihood is the original definition.
  \item It is used, and understood, by many thousands of applied mathematicians.
  \item It has been in use for over 80 years.
  \item The MW definition is a restriction of the original definition. 
  \item It is only used, and understood, by a very small number of meteorologists involved in the field
  of probabilistic forecast verification.
  \item It is very recent.
\end{itemize}

\section{Summary}
\label{conclusions}

Statisticians have used the phrase \emph{likelihood} for over 80 years, with a particular meaning, 
following~\citet{fisher1922}.
A relatively recent paper by~\citet{murphyw87} attempts to redefine this word 
when applied to meteorological forecasts and observations.
This undermines the original meaning, creates confusion, and is not very helpful in 
building connections between statistics and meteorology.
We therefore strongly advise that meteorologists working in forecast verification should \emph{not} use
the definition of~\citet{murphyw87}, and should stick to the original definition of~\cite{fisher1922}.

\section{Acknowledgements}

SJ would like to thank Beth Ebert for helpful discussions (although she doesn't necessarily agree with the views
expressed in the article).

\bibliographystyle{plainnat}
\bibliography{noteonlikelihood}

\begin{thebibliography}{7}
\expandafter\ifx\csname natexlab\endcsname\relax\def\natexlab#1{#1}\fi
\expandafter\ifx\csname url\endcsname\relax
  \def\url#1{{\tt #1}}\fi

\bibitem[Casella and Berger(2002)]{casellab02}
G~Casella and R~L Berger.
\newblock {\em Statistical Inference}.
\newblock Duxbury, 2002.

\bibitem[Fisher(1922)]{fisher1922}
R~Fisher.
\newblock On the mathematical foundations of statistics.
\newblock {\em Philosophical Transactions of the Royal Society, A},
  222:\penalty0 309--368, 1922.

\bibitem[Jewson et~al.(2003)Jewson, Brix, and Ziehmann]{jewsonbz03a}
S~Jewson, A~Brix, and C~Ziehmann.
\newblock A new framework for the assessment and calibration of ensemble
  temperature forecasts.
\newblock {\em ASL}, 2003.
\newblock Submitted.

\bibitem[Jolliffe and Stephenson(2003)]{jolliffes03}
I~Jolliffe and D~Stephenson.
\newblock {\em Forecast Verification: A Practioner's Guide in Atmospheric
  Science}.
\newblock Wiley, 2003.

\bibitem[Murphy and Winkler(1987)]{murphyw87}
J~Murphy and R~Winkler.
\newblock A general framework for forecast verification.
\newblock {\em Monthly Weather Review}, 115:\penalty0 1330--1338, 1987.

\bibitem[Press et~al.(1992)Press, Teukolsky, Vetterling, and Flannery]{nr}
W~Press, S~Teukolsky, W~Vetterling, and B~Flannery.
\newblock {\em Numerical Recipes}.
\newblock Cambridge University Press, 1992.

\bibitem[Wilks(2001)]{wilks95}
D~Wilks.
\newblock {\em Statistical methods in the atmospheric sciences}.
\newblock Academic Press, 2001.

\end{thebibliography}

\end{document}